\documentclass[10pt,final,conference,twocolumn,a4paper]{IEEEtran}

\usepackage{amsmath,amssymb,amsfonts}
\usepackage{algorithmic}
\usepackage{graphicx}
\usepackage{textcomp}
\usepackage{xcolor}
\usepackage{booktabs}
\usepackage{cite}
\usepackage{url}
\usepackage{bm}
\usepackage{multirow}

\usepackage[capitalize]{cleveref}

\IEEEoverridecommandlockouts

\DeclareMathOperator{\DFT}{DFT}
\DeclareMathOperator{\IDFT}{IDFT}
\newcommand{\Hest}{\hat{\mathbf{H}}}
\newcommand{\Htrue}{\mathbf{H}}
\newcommand{\Hmask}{\mathbf{H}_{\mathrm{m}}}

\newcommand{\CIR}{\mathbf{h}}
\newcommand{\cest}{\hat{\mathbf{h}}}
\newcommand{\real}{\mathbb{R}}
\newcommand{\complex}{\mathbb{C}}
\newcommand{\dmodel}{d_{\mathrm{model}}}
\usepackage{todonotes}

\RequirePackage{xstring}
\RequirePackage{xparse}
\RequirePackage{acro}
\makeatletter
\@ifpackagelater{acro}{2019/02/17}{
  \NewDocumentCommand\acrodef{mO{#1}mG{}}{\DeclareAcronym{#1}{short={#2}, long={#3}, foreign-plural={}, #4}}
}{
  \NewDocumentCommand\acrodef{mO{#1}mG{}}{\DeclareAcronym{#1}{short={#2}, long={#3}, #4}}
}
\makeatother

\DeclareAcronym{CFR}{short=CFR, short-plural=, long=channel frequency response}
\DeclareAcronym{CNN}{short=CNN, short-plural=,long=convolutional neural
networks}
\DeclareAcronym{RNN}{short=RNN, short-plural=,long=recurrent networks}
\DeclareAcronym{PDP}{short=PDP, short-plural=,long=power delay profile}
\DeclareAcronym{CIR}{short=CIR, short-plural=,long=channel impulse response}
\DeclareAcronym{DTMC}{short=DTMC, short-plural=, long= Discrete-Time Markov Chain}
\DeclareAcronym{FNN}{short=FNN, short-plural=, long=Feed-Forward Network}
\DeclareAcronym{GELU}{short=GELU, short-plural=, long=Gaussian Error Linear Unit}
\DeclareAcronym{MSE}{short=MSE, short-plural=, long=mean squared error}
\DeclareAcronym{UE}{short=UE, short-plural=, long=user equipment}
\DeclareAcronym{LPF}{short=LPF, short-plural=, long=low pass filter}
\DeclareAcronym{LSTM}{short=LSTM, short-plural=, long=long short-term memory}
\DeclareAcronym{GRU}{short=GRU, short-plural=, long=gated recurrent unit}
\DeclareAcronym{OFDM}{short=OFDM, short-plural=, long=orthogonal frequency-division multiplexing}
\DeclareAcronym{CSI}{short=CSI, short-plural=, long=channel state information}
\begin{document}

\title{Physics-Informed Transformer for
Multi-Band Channel Frequency Response Reconstruction 
}

\author{
    \IEEEauthorblockN
    {%
    Anatolij Zubow, Joana Angjo, Sigrid Dimce, and Falko Dressler
    }%
    \IEEEauthorblockA
    {%
        School of Electrical Engineering and Computer Science, TU Berlin, Berlin, Germany\\
        \{zubow, angjo, dimce, dressler\}@ccs-labs.org
    }%
}

\maketitle

\begin{abstract}
Wideband \ac{CFR} estimation is challenging in multi-band wireless systems, especially when one or more sub-bands are temporarily blocked by co-channel interference.
We present a physics-informed complex Transformer that reconstructs the full wideband \ac{CFR} from such fragmented, partially observed spectrum snapshots.
The interference pattern in each sub-band is modeled as an independent two-state discrete-time Markov chain, capturing realistic bursty occupancy behavior.
Our model operates on the joint time--frequency grid of $T$ snapshots and $F$ frequency bins and uses a factored self-attention mechanism that separately attends along both axes, reducing the computational complexity to $\mathcal{O}(TF^2 + FT^2)$.
Complex-valued inputs and outputs are processed through a holomorphic linear layer that preserves phase relationships.
Training uses a composite physics-informed loss combining spectral fidelity, \ac{PDP} reconstruction, \ac{CIR} sparsity, and temporal smoothness.
Mobility effects are incorporated through per-sample velocity randomization, enabling generalization across different mobility regimes.
Evaluation against three classical baselines, namely, last-observation-carry-forward, zero-fill, and cubic-spline interpolation, shows that our approach achieves the highest \ac{PDP} similarity with respect to the ground truth, reaching $\rho \geq 0.82$ compared to $\rho \geq 0.62$ for the best baseline at interference occupancy levels up to 50\%.
Furthermore, the model degrades smoothly across the full velocity range 
consistently outperforming all other baselines.
\end{abstract}


\section{Introduction}
\label{sec:intro}
Wireless sensing applications such as passive user detection, indoor localization, and gesture recognition in 5G NR and IEEE 802.11bf rely on accurate wideband \ac{CFR} estimates~\cite{wei2022toward}. 
The \ac{CFR} encodes the multipath delays, Doppler shifts, and scatterer geometry, from which sensing features are extracted. 
In practice, co-channel interference from neighboring cells, unlicensed devices, or radars can block entire sub-bands for multiple consecutive time slots, leaving an incomplete time--frequency grid and degrading the delay and Doppler resolution on which sensing depends. 
Reconstructing the full wideband \ac{CFR} from fragmented observations is thus of direct practical importance in integrated sensing and communication systems.

%
%
%
%

Classical gap-filling strategies like zero-filling, last-observation carried-forward, and frequency-domain spline interpolation are computationally inexpensive, but make no use of the structure of the radio channel (multipath, fading, correlation between subcarriers over time).
They are particularly fallible when sub-bands are missing/occupied for multiple time slots, since there is no spectral information from which to correctly infer the missing values.
Current data-driven methods, such as \ac{CNN}, are promising, but they have a key weakness: they cannot easily link information across distant frequency subcarriers. 
Because they lack a global attention mechanism, they miss out on wideband coherence and frequently fail to model how the channel evolves over time.
Transformer architectures on the other hand have shown strong generalization capabilities in sequence modeling tasks~\cite{vaswani2017attention}, and their usage for channel estimation is increasing~\cite{he2020modeldriven}. 
However, two characteristics of wireless channels make a direct application of standard Transformers suboptimal. 
First, the channel is \textit{complex-valued}, and simply treating the real and imaginary parts as independent ignores the intrinsic relationships that encode phase information and propagation delays. 
Second, the channel evolves over a \textit{two-dimensional time-frequency grid} ($T\,\times\,F$); and flattening it into a 1D sequence makes the computation scale quadratically with $TF$ limiting its use in wideband systems.

Motivated by this, an architecture denoted as \emph{CFRTransformer} is presented in this paper.
The main contributions are:
\begin{itemize}
  \item We propose a factored self-attention mechanism that alternates between frequency (i.e., learning spectral correlations across bands) and time domain (i.e., learning Doppler-coherent temporal dynamics). 
  \item We use a holomorphic \emph{ComplexLinear} embedding layer that maps the complex-valued input features to the model dimension, preserving phase relationships. 
  \item We employ a positional encoding that injects absolute bin-position information along the frequency axis. 
  \item We develop a multi-objective loss function that incorporates domain-specific constraints. 
  By penalizing errors in the \ac{PDP}, promoting \ac{CIR} sparsity, and enforcing temporal smoothness, the model produces estimates consistent with the channel coherence time, and ensures that the training loss is consistent with physics.
  \item We perform mobility-randomized training over velocities $v \in [0.5,\,30]$~m/s, and evaluate reconstruction quality across the full mobility range. 
\end{itemize}

\section{System Model}
\label{sec:sysmodel}

\subsection{Multi-Band Wideband Channel}
\label{ssec:channel}
We consider a wideband channel observed over $N_\mathrm{b}$ non-overlapping sub-bands, each comprising of $F_\mathrm{b}$ frequency bins, giving $F = N_\mathrm{b} F_\mathrm{b}$ bins spanning the system bandwidth $B$.
The channel is sampled over $T$ consecutive snapshots of duration $T_\mathrm{s}$, producing a complex matrix $\Htrue \in \complex^{T \times F}$ where $H[t,f]$ is the complex channel gain at snapshot $t$ and frequency bin $f$.
The channel impulse response (CIR) at snapshot $t$ is modeled as the sum of $P$ discrete multipath components:
\begin{equation}
  h[t,n] = \sum_{p=1}^{P} g_p(t)\,\delta\!\left[n - d_p(t)\right],
  \label{eq:cir}
  \vspace{-0.5em}
\end{equation}
where $d_p(t) \in \{0,\ldots,F{-}1\}$ is the delay tap of path $p$ at snapshot $t$, and $g_p(t) \in \complex$ is its complex gain. 
The CFR is obtained as
$H[t,:] = \DFT\bigl(h[t,:]\bigr)$, where DFT denotes the discrete Fourier transform operation.
To capture user mobility, the channel evolves over time, inducing Doppler shifts in the multipath components.
The complex gain evolves as:
\begin{equation}
  g_p(t) = |g_p|\,e^{j(\varphi_{p,0} + \Delta\varphi_p t)}
           + 0.1\bigl(n_{r,p}(t) + jn_{i,p}(t)\bigr),
  \label{eq:gain}
\end{equation}
where $|g_p|^2 \sim \mathrm{Exp}(1)$ (Rayleigh envelope), $\varphi_{p,0} \sim \mathcal{U}[0,2\pi)$ is a random initial phase, and $\Delta\varphi_p \sim \mathcal{U}(-\Delta\varphi_\mathrm{max},
+\Delta\varphi_\mathrm{max})$ is the per-snapshot Doppler phase increment with
$\Delta\varphi_\mathrm{max} = 2\pi f_{d,\mathrm{max}} T_\mathrm{s}$.
The maximum Doppler shift is $f_{d,\mathrm{max}} = v_\mathrm{max} f_c / c$, where $f_c$ is the carrier frequency and $c$ is the speed of light.
The noise terms $n_{r,p}(t)$, $n_{i,p}(t)$ are obtained by passing independent white Gaussian samples through a second-order Butterworth low-pass filter with normalized cutoff $W_n = f_{d,\mathrm{max}} T_\mathrm{s}/0.423$, modeling band-limited time-varying fading consistent with Clarke's isotropic scattering model~\cite{clarke1968statistical}.
%
%
It is multiplied by a factor of 0.1 so that it contributes about 1\% of the average path power, ensuring that it represents mild gain fluctuations, consistent with Clarke's model~\cite{clarke1968statistical}, while keeping the main Doppler phase behavior of each path intact.
Finally the delay tap $d_p$ is drawn uniformly from $\{0,\ldots,d_{\max}\}$ and perturbed by $\pm 1$-bin jitter per snapshot. 
This approximates sub-bin timing uncertainty arising from the discrete delay grid of a DFT-based representation of resolution $\Delta\tau = 1/B$, and prevents the model from exploiting fixed tap positions as a reconstruction shortcut during
training.
\subsection{\ac{DTMC} Interference Model}
\label{ssec:dtmc}

Each sub-band $b$ is modeled by a two-state \ac{DTMC} with states $\mathcal{S} = \{\mathrm{Idle}, \mathrm{Busy}\}$ via a binary mask $M[t,f]$, where $0$ indicates idle (CFR observed) and $1$ busy (CFR missing).  
The masked observation is:
\begin{equation}
  \Hmask[t,f] = H[t,f] \,(1-M[t,f]).
  \label{eq:mask}
\end{equation}
The transition probabilities are
$p_{01} = P(\mathrm{Idle}{\to}\mathrm{Busy})$ and
$p_{10}$ vice-versa.
Given a target stationary busy probability $\pi_\mathrm{busy}$,
\begin{equation}
  p_{01} = \frac{\pi_\mathrm{busy}\,p_{10}}{1-\pi_\mathrm{busy}},
  \qquad
  \pi_\mathrm{busy} = \frac{p_{01}}{p_{01}+p_{10}}.
  \label{eq:dtmc}
\end{equation}
The chain is initialized by sampling the state at $t{=}0$ from the stationary distribution.
With $p_{10}{=}0.30$, the mean burst duration is $1/p_{10} \approx3.3$ snapshots, consistent with realistic co-channel interference bursts.
%
%
While sub-bands are modeled independently, assuming narrow-band interference and neglecting potential ``vertical'' frequency correlations, this remains a robust first-order approximation for most co-channel scenarios.

\section{CFRTransformer}
\label{sec:method}

\subsection{Input Features and ComplexLinear Embedding}
\label{ssec:embed}

The network receives three real-valued feature channels per time--frequency grid node $(t,f)$:
\begin{equation}
  \mathbf{x}[t,f] =
  \bigl[\mathrm{Re}(\Hmask),\,\mathrm{Im}(\Hmask),\,
        M\bigr],
  \label{eq:features}
  \vspace{-0.7em}
\end{equation}
where $\mathrm{Re}(\Hmask)$ and $\mathrm{Im}(\Hmask)$ are is the real and imaginary parts of the masked \ac{CFR}, and $M$ is the binary mask indicating sub-band occupancy. 
%
%
The features are embedded to dimension $\dmodel$ via a \emph{ComplexLinear} layer~\cite{trabelsi2017deep}, which processes the real part $\mathbf{x}_r$ and imaginary part $\mathbf{x}_i$ jointly:
\begin{equation}
  \mathrm{out}_r = \mathbf{W}_r \mathbf{x}_r - \mathbf{W}_i \mathbf{x}_i,
  \qquad
  \mathrm{out}_i = \mathbf{W}_i \mathbf{x}_r + \mathbf{W}_r \mathbf{x}_i.
  \label{eq:complexlinear}
    \vspace{-0.7em}
\end{equation}
The shared weight matrices $\mathbf{W}_r, \mathbf{W}_i$ satisfy the Cauchy-Riemann conditions, making the mapping holomorphic and ensuring that complex phase relationships are preserved.
The output head also uses a layer of the same design, mapping the model features back to the complex \ac{CFR} space.
\subsection{Frequency Positional Encoding}
Since attention layers are agnostic to the order of their inputs, they cannot naturally discern the relative positions of frequency bins. 
We therefore incorporate sinusoidal positional encoding~\cite{vaswani2017attention} along the frequency axis to provide the model with absolute spatial context. 
This step is critical; without it, the bins would be treated as an unordered set, making accurate reconstruction impossible.
%
%
%
%
The encoding is computed as:
\begin{align}
  \mathrm{pe}[f, 2k]   &= \sin\!\bigl(2\pi \tfrac{f}{F-1}(k+1)\bigr), \\
  \mathrm{pe}[f, 2k+1] &= \cos\!\bigl(2\pi \tfrac{f}{F-1}(k+1)\bigr),
  \vspace{-0.8em}
\end{align}
where $\mathrm{pe}[f,:]$ denotes the positional encoding vector for frequency bin $f \in \{0,\ldots,F{-}1\}$ and $k = 0,1,\ldots,d_{\mathrm{model}}/2{-}1$.
This is added to both the real and imaginary tensors.
Because the $N_b$ sub-bands are non-overlapping and contiguous, this way every frequency bin has a globally unique index $f$. 
%

%
%
%
\subsection{Factored Attention Blocks}
\label{ssec:attention}
Flattening the $T{\times}F$ grid into a single sequence of length $TF$ would make self-attention $\mathcal{O}((TF)^2)$, which is intractable for practical configurations. 
Instead, each \emph{FactoredAttentionBlock} applies attention along two axes sequentially:

\smallskip

\noindent \subsubsection{Frequency attention}
The tensor $\mathbf{X} \in \real^{B \times T \times F \times \dmodel}$, representing a batch of embedded time-frequency grids, is reshaped to $(BT, F, \dmodel)$, treating each time snapshot as an independent sequence of $F$ tokens.
Multi-head attention ($N_h{=}4$ heads) is applied to capture dependencies between bins, followed by a residual connection and layer normalization.
This pass learns spectral correlations such as inter-band coherence, transitions at band edges, and the gradual spectral variations imposed by the multipath propagation.

\smallskip
\noindent \subsubsection{Time attention}
To complement the previous, attention is then applied along the time axis to capture temporal dynamics.
The tensor is permuted and reshaped to $(BF, T, \dmodel)$, treating each frequency bin as an independent sequence of $T$ tokens.
A second multi-head attention is applied with residual and LayerNorm. 
This pass learns temporal dynamics: Doppler-induced phase ramps, interference burst patterns, and channel coherence across snapshots.

The complexity is reduced to $\mathcal{O}(TF^2+FT^2)$, saving a factor of $(TF)/(T{+}F)$ which is ${\approx}\,20{\times}$ for a grid with $T{=}20$ and $F{=}1280$. 
After $N_\mathrm{b}{=}2$ stacked blocks, a position-wise \ac{FNN} (Linear$(\dmodel,2\dmodel)$\,-\,\ac{GELU}\,-\,Linear$(2\dmodel,\dmodel)$) with a residual connection and LayerNorm produces the final feature vector, which is projected to a complex scalar per grid node by the ComplexLinear output head.

\subsection{Physics-Informed Training Loss}
\label{ssec:loss}
To train the network effectively while complying with the physical properties of a wireless channel, the total loss combines four terms, as follows:
\begin{equation}
  \mathcal{L} = \mathcal{L}_\mathrm{CFR}
              + \lambda_\mathrm{pdp}\,\mathcal{L}_\mathrm{PDP}
              + \lambda_\mathrm{sp}\,\mathcal{L}_\mathrm{sparse}
              + \lambda_\mathrm{t}\,\mathcal{L}_\mathrm{temp}.
  \label{eq:loss}
\end{equation}
\noindent \subsubsection{Spectral fidelity}
%
\begin{equation}
  \mathcal{L}_\mathrm{CFR} = \mathbb{E}\bigl[\lvert\Hest - \Htrue\rvert^2\bigr],
  \label{eq:lcfr}
\end{equation}
is the \ac{MSE} between the estimated \ac{CFR} $\hat{H}$ and the true \ac{CFR} $H$, encouraging the correct reconstruction of both amplitude and phase.
%

\smallskip

\noindent \subsubsection{\ac{PDP} fidelity}
\begin{equation}
  \mathcal{L}_\mathrm{PDP} =
  \mathbb{E}\!\left[\Bigl(\lvert\cest\rvert^2
                  - \lvert\CIR\rvert^2\Bigr)^2\right],
  \label{eq:lpdp}
\end{equation}
is the \ac{MSE} between $\cest = \IDFT(\Hest)$ and $\CIR = \IDFT(\Htrue)$.
This term penalizes errors in the \ac{PDP} directly, correcting for phase ambiguity in $\mathcal{L}_\mathrm{CFR}$.
The unit-normalized \ac{PDP} and its corresponding estimate are denoted by $\bm{p}$ and $\hat{\bm{p}}$.
The \ac{PDP} similarity factor, computed as follows \cite{xie2019precise}, is used for evaluation:
\begin{equation}
  \rho = 1 - \frac{\|\bm{\hat{p}} - \bm{p}\|_2}{\sqrt{2}}.
  \label{eq:rho}
\end{equation}
Here, the $\sqrt{2}$ normalization ensures $\rho \in [0,1]$ with $\rho{=}1$ for a perfect match and $\rho{=}0$ for maximum dissimilarity.
\smallskip

\noindent \subsubsection{CIR sparsity}
%
\begin{equation}
\mathcal{L}_\mathrm{sparse} = \mathbb{E}[\lvert\cest\rvert],
  \label{eq:lsparse}
\end{equation}
is the mean absolute value (L1 regularization) of the estimated \ac{CIR} amplitudes. 
This takes into account the physical assumption that real channels have
only a few dominant propagation taps (e.g., $P=6$) out of $F$ possible intervals, thus avoiding non-realistic solutions.
This is reasonable since most received energy in wireless channels is concentrated in the earliest paths~\cite{tse2005fundamentals}.
Standardized LTE and 5G models also use tapped delay lines with 7-9 taps~\cite{3gpp38901}.

\smallskip

\noindent \subsubsection{Temporal smoothness}
\begin{equation}
  \mathcal{L}_\mathrm{temp} =
  \mathbb{E}\!\left[\bigl\lvert
    \Hest[:,t{+}1] - \Hest[:,t]
  \bigr\rvert\right], \ t=1 \ldots T{-}1,
  \label{eq:ltemp}
\end{equation}
is computed only on the estimated \ac{CFR}.
It penalizes large frame-to-frame fluctuations, encouraging temporal consistency imposed by the channel coherence time. 


\subsection{Velocity-Randomized Training}
\label{ssec:training}

A critical design choice is that the \ac{UE} velocity is randomized \emph{per training sample} as $v \sim \mathcal{U}(0.5, 30)$~m/s.
This randomizes the Doppler spread, the channel coherence time, and the low pass filter bandwidth used to generate the correlated gain noise for each training instance.
Without this variability, the temporal attention would learn patterns suited only to a single Doppler regime, and would lead to  miscalibrations for other velocities.
The occupancy is also drawn independently from $\pi_\mathrm{busy} \sim \mathcal{U}(0.1, 0.9)$, exposing the model to a wide range of conditions.
%

%
%
%
\subsection{Hyperparameter Selection}

The loss weights $(\lambda_{\mathrm{pdp}}, \lambda_{\mathrm{sp}}, \lambda_{\mathrm{t}}) = (1.0,\; 5\times10^{-4},\; 0.05)$ were chosen such that gradients are of comparable magnitude at initialization. 
The small $\lambda_{\mathrm{sp}}$ accounts for the lower magnitude of \ac{CIR} taps compared to \ac{CFR}, while $\lambda_{\mathrm{t}}$ avoids over-penalizing Doppler-induced phase rotations.
%
%
Training uses AdamW with learning rate $10^{-3}$, weight decay $10^{-4}$, cosine annealing over 70 epochs, and gradient clipping at $\ell_2$-norm $=1$ \cite{vaswani2017attention}.
Each epoch consists of 5{,}000 steps with one sample per step. 
Per-sample randomization of $v \sim \mathcal{U}(0.5, 30)$~m/s and $\pi_{\mathrm{busy}} \sim \mathcal{U}(0.1, 0.9)$ ensures coverage of diverse operating conditions.

\section{Evaluation Methodology}
\label{sec:eval}
The results report the value of the \ac{PDP} similarity factor (\cref{eq:rho}) that corresponds to the mean across all snapshots.

\subsection{Baselines}


\textbf{A – Historical fill:}
Each missing bin is replaced by the last observed value at that frequency bin, scanning forward in time.
This approach works well at low Doppler, but not when the channel varies faster than the interference burst duration. 

\textbf{B – Zero-fill:}
Here, missing bins are filled with zeros.
This trivial approach introduces spectral discontinuities, which spread energy across all delay taps in the \ac{PDP}.

\textbf{C – Cubic-spline frequency interpolation:}
A 1D cubic spline is fitted to the observed bins at each snapshot and evaluated at the missing positions.
Real and imaginary parts are interpolated independently.
When fewer than four observed bins are available, the method falls back to linear interpolation.

\subsection{Evaluation Axes}

Performance is evaluated along two independent sweeps:

\noindent \textit{Occupancy sweep:}
The interference busy probability $\pi_\mathrm{busy}$ is swept over $\{0.1, 0.3, 0.5, 0.7, 0.9\}$ at a fixed velocity of $v{=}7$~m/s.
This replicates the standard evaluation found in spectrum-sharing studies and shows whether performance degrades gradually and predictably as interference increases.

\noindent \textit{Velocity sweep:}
The velocity is swept over $\{0.5, 1, 3, 7, 15, 30\}$~m/s at fixed $\pi_\mathrm{busy}{=}0.5$ (mid-load interference).
This axis, absent from most prior channel-reconstruction papers, evaluates generalization across full mobility range, i.e. from pedestrian to motorway mobility regimes.
The remaining parameters are summarized in Table~\ref{tab:params}.

\begin{table}[t]
\caption{Simulation Parameters}
\vspace{-0.6em}
\label{tab:params}
\centering
\renewcommand{\arraystretch}{1.1}
\begin{tabular}{lll}
\toprule
\textbf{Parameter} & \textbf{Symbol} & \textbf{Value} \\
\midrule
Carrier frequency      & $f_c$           & 3.5~GHz \\
Total bandwidth        & $B$             & 100~MHz \\
Number of sub-bands    & $N_\mathrm{b}$  & 5 (each 20~MHz)\\
Bins per sub-band      & $F_\mathrm{b}$  & 256 \\
Total frequency bins   & $F$             & 1280 \\
Freq. resolution       & $\Delta f$      & 78.1~kHz/bin \\
Snapshot duration      & $T_\mathrm{s}$  & 0.5~ms (=LTE slot) \\
Number of snapshots    & $T$             & 20 \\
Number of paths        & $P$             & 6 (2, 10) \\
\midrule
Model dimension        & $\dmodel$       & 128 \\
Attention heads/blocks & $N_h/N_b$           & 4 / 2 \\
FFN hidden dimension         & ---             & 256 \\
\midrule
Training velocity range & $[v_\mathrm{min},v_\mathrm{max}]$ & [0.5, 30]~m/s \\
Occupancy range (train) & $\pi_\mathrm{busy}$ & $\mathcal{U}(0.1,0.9)$ \\
Training samples / epoch & --- & 5000 \\
Epochs                  & ---  & 70 \\
\midrule
Test samples per point  & ---  & 500 \\
Occupancy levels (eval.) & ---  & \{0.1, 0.3, 0.5, 0.7, 0.9\} \\
Velocity levels (eval.)  & ---  & \{0.5, 1, 3, 7, 15, 30\}~m/s \\
Fixed occ.\ (vel.\ eval.) & --- & 0.5 \\
\bottomrule
\end{tabular}
\end{table}

\section{Results and Discussion}
\label{sec:results}

\subsection{Qualitative \ac{CFR} and \ac{PDP} Reconstruction}
\begin{figure*}[t]
  \centering
  \includegraphics[width=0.65\linewidth]{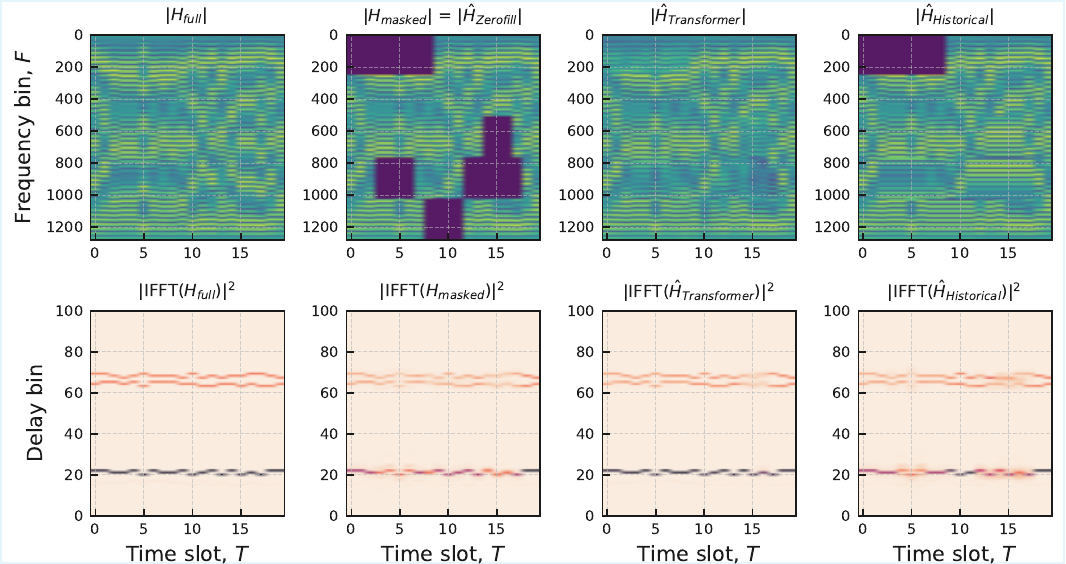}
  \caption{CFR magnitude (top) and PDP (bottom) for a single representative trace at 30\% interference occupancy.
  From left to right: clean full-band reference $H_{\mathrm{full}}$; masked observation $H_{\mathrm{masked}}$ (equivalent to zero-fill); \emph{CFRTransformer} reconstruction $\hat{H}_{\mathrm{Transformer}}$; historical-fill reconstruction $\hat{H}_{\mathrm{Historical}}$.
  Dark horizontal bands in the masked CFR indicate blocked sub-bands.
  }
  \label{fig:paper_eval_recon_0}
      \vspace{-1.1em}
\end{figure*}

Fig.~\ref{fig:paper_eval_recon_0} shows an example of the \ac{CFR} magnitude and the corresponding \ac{PDP} for four cases at 30\% interference occupancy: the clean full-band reference, the zero-filled masked observation, the \emph{CFRTransformer} reconstruction, and the historical-fill reconstruction.
In the \ac{PDP} plots, the clean reference shows sharp and well-localized delay peaks, whereas the outputs of historical fill and zero-fill (Strategy A and B) show visible smearing and energy spread across delay bins, distortions that are
substantially reduced by \emph{CFRTransformer}.

\subsection{PDP Similarity vs.\ Interference Occupancy}
\label{ssec:res_occ}

\begin{figure}[t]
  \centering
  \includegraphics[width=0.67\linewidth]{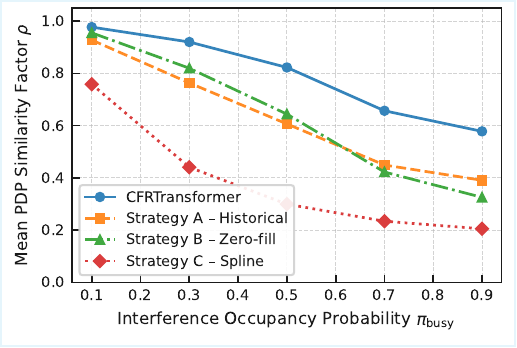}
  \vspace{-0.8em}
  \caption{Mean PDP similarity factor $\rho$ vs.\ interference
        probability $\pi_\mathrm{busy}$.
           }
  \label{fig:rho_vs_occ}
\end{figure}

Fig.~\ref{fig:rho_vs_occ} shows the mean $\rho$ as a function of $\pi_\mathrm{busy}$ for all strategies.
\emph{CFRTransformer} achieves the highest $\rho$ at every occupancy level.
At low occupancy ($\pi_\mathrm{busy}{=}0.1$), all methods perform well, since most bins are observed.
However, at $\pi_\mathrm{busy}{=}0.5$, while \emph{CFRTransformer} maintains $\rho{\approx}0.82$, cubic-spline drops to ${\approx}0.30$ and historical fill to ${\approx}0.61$.
Zero-fill falls below~0.5 at $\approx\pi_\mathrm{busy}{=}0.63$ due to the spectral discontinuities.

At high occupancy ($\pi_\mathrm{busy}{=}0.9$), only one sub-band in ten is typically idle at any snapshot.
\emph{CFRTransformer} still achieves $\rho{\approx}0.58$, leveraging the temporal context across all $T=20$ snapshots and the cross-band continuity captured by the attention weights.
On the other hand, the spline and zero-fill strategies degrade to $\rho{\approx}0.33$ and $\rho{\approx}0.21$.
The monotonic decrease of $\rho$ with $\pi_\mathrm{busy}$ for all methods confirms that the \ac{DTMC} mask model provides a well-graded difficulty axis.
%

%
\subsection{\ac{PDP} Similarity vs.\ \ac{UE} Velocity}
\label{ssec:res_vel}

\begin{figure}[t]
  \centering
  \includegraphics[width=0.75\linewidth]
  {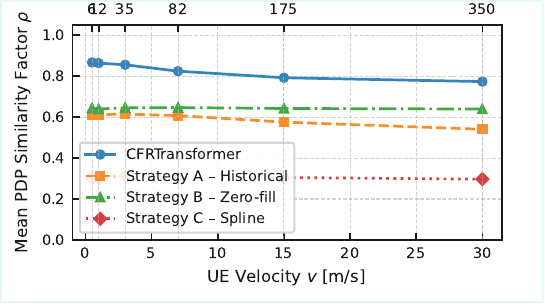}
  \vspace{-0.8em}
  \caption{Mean PDP similarity factor $\rho$ vs.\ UE velocity at
           fixed $\pi_\mathrm{busy}{=}0.5$. The secondary axis shows maximum Doppler shift $f_d$~[Hz]. 
           }
  \label{fig:rho_vs_vel}
  \vspace{-.8em}
\end{figure}

Fig.~\ref{fig:rho_vs_vel} presents the novel velocity-sweep evaluation at fixed $\pi_\mathrm{busy}{=}0.5$.
The x-axis shows velocity in m/s and the secondary axis at the top shows the corresponding maximum Doppler shift $f_d = v f_c / c$ in Hz.
\emph{CFRTransformer} achieves $\rho{\approx}0.87$ at 0.5~m/s (quasi-static indoor, $f_d{\approx}6$~Hz) and degrades smoothly to $\rho{\approx}0.79$ at 30~m/s (fast vehicular, $f_d{\approx}350$~Hz).
This slight drop over a $60\times$ increase in velocity confirms that the velocity-randomized training effectively generalizes across the mobility range.
Strategy~A is the most mobility-sensitive method, achieving moderate performance at 0.5~m/s ($\rho{\approx}0.61$) but degrading to $\rho{\approx}0.54$ at 30~m/s, where a single-snapshot carry-forward is phase-incoherent with the true channel.
Cubic spline has the poorest performance of all, but it is less velocity-sensitive as it operates per-snapshot and does not exploit temporal information.
Zero-fill gives the second highest performance and it is also insensitive to velocity, since it performs no interpolation.
\emph{CFRTransformer} consistently outperforms all of them, as its time-attention mechanism adapts to the shifting temporal correlation structure.
\subsection{Impact of Channel Complexity on Velocity Robustness}

\begin{figure}[t]
  \centering
  \includegraphics[width=0.67\linewidth]{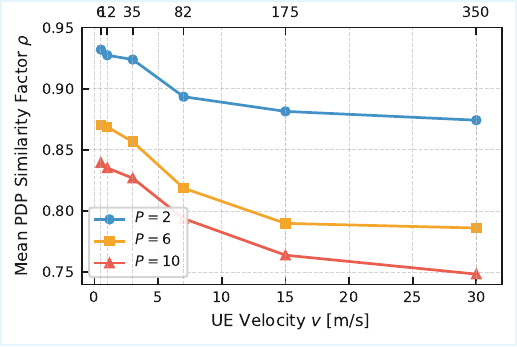}
  \vspace{-0.6em}
  \caption{Mean PDP similarity factor $\rho$ vs.\ UE velocity evaluated at three channel complexities: $P \in \{2, 6, 10\}$ multipath components ($\pi_{\mathrm{busy}}=0.5$). 
  }
  \label{fig:rho_vs_vel_paths}
\end{figure}

Fig.~\ref{fig:rho_vs_vel_paths} shows the mean $\rho$ of the \emph{CFRTransformer} 
as a function of \ac{UE} velocity and three different channel complexities, namely consisting of $P \in \{2, 6, 10\}$ multipath components, at fixed $\pi_{\mathrm{busy}}=0.5$.
As expected, $\rho$ decreases with both velocity and $P$: a sparse channel ($P=2$) remains easy to reconstruct ($\rho \geq 0.875$), while a richer channel ($P=10$) degrades from $\rho \approx 0.84$ at low velocity to $\rho \approx 0.75$ at $v=30$~m/s.
In all cases, \emph{CFRTransformer} degrades smoothly with velocity.
The steepest drop occurs below $v \approx 15$~m/s, after which the mean $\rho$ nearly saturates, showing that the model has learned channel patterns that remain accurate even at high Doppler speeds.

\subsection{Impact of Number of Sub-Bands on Reconstruction Quality}
\begin{figure}[t]
  \centering
  \includegraphics[width=0.67\columnwidth]{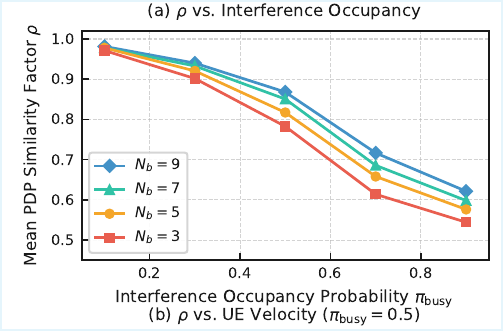}
  \vspace{-0.6em}
  \includegraphics[width=0.67\columnwidth]{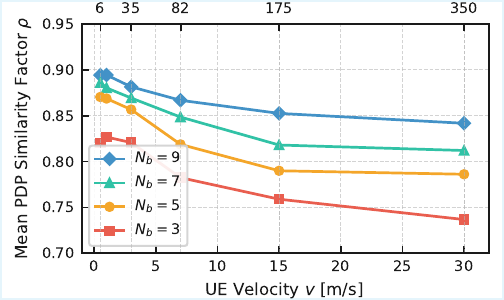}
  \caption{Mean PDP similarity factor $\rho$ for \emph{CFRTransformer} as a function of (a) interference occupancy $\pi_{\mathrm{busy}}$ at $v=7$~m/s, and (b) UE velocity at $\pi_{\mathrm{busy}}=0.5$, for $N_b \in \{3, 5, 7, 9\}$ sub-bands.}
  \vspace{-0.8em}
  \label{fig:rho_vs_bands}
\end{figure}

Fig.~\ref{fig:rho_vs_bands} shows mean $\rho$ as a function of interference occupancy (up) and \ac{UE} velocity (down) for \emph{CFRTransformer} trained and evaluated with $N_b \in \{3, 5, 7, 9\}$ sub-bands at a fixed total bandwidth.
Performance improves with increasing $N_b$: at $\pi_{\mathrm{busy}}=0.5$, $\rho$ rises from $0.78$ ($N_b=3$) to $0.87$ ($N_b=9$), as wider spectral context leads to better gap filling.
This trend holds across velocities, with $N_b=9$ maintaining $\rho \geq 0.84$ at $v=30$~m/s versus $\rho \approx 0.74$ for $N_b=3$.

\subsection{Effect of Velocity Randomization}

To quantify the benefit of velocity-randomized training, Table~\ref{tab:ablation} compares models trained at a fixed velocity, $v \in \{0.5, 7, 30\}$~m/s, against the full velocity-randomized model, both evaluated with $\pi_\mathrm{busy}{=}0.5$.
The fixed-velocity models show significant performance degradation when tested at a different velocity. 
A model trained at a fixed velocity of $v = 0.5$~m/s drops significantly in performance ($\Delta\rho \approx 0.20$) at high velocity.
The temporal smoothness weight $\lambda_\mathrm{t}$, tuned for 0.5~m/s, becomes too aggressive at 30~m/s, over-smoothing the rapidly varying \ac{CFR}.
On the other hand, velocity randomization removes this calibration dependency thus performance similar to that of fixed-velocity models is maintained.
\begin{table}[t]
\caption{Ablation: Fixed vs.\ Randomized Velocity Training
\vspace{-0.6em}
         ($\pi_\mathrm{busy}{=}0.5$)}
\label{tab:ablation}
\centering
\renewcommand{\arraystretch}{1.1}
\begin{tabular}{lccc}
\toprule
\textbf{Training condition} & $\rho$ at 0.5~m/s & $\rho$ at 7~m/s & $\rho$ at 30~m/s \\
\midrule
Fixed $v{=}0.5$~m/s           & \textbf{0.886}             & 0.765 &     0.686 \\
Fixed $v{=}7$~m/s           & 0.872             & \textbf{0.825} &       0.775\\
Fixed $v{=}30$~m/s           & 0.835             & 0.804 &       0.770\\
Random $v \sim \mathcal{U}(0.5,30)$~m/s
                             & 0.870    & 0.819 &     \textbf{0.786} \\
\bottomrule
\end{tabular}
\end{table}
%

\subsection{Computational Overhead}
At inference, \emph{CFRTransformer} processes one $(T,F)=(20,1280)$ grid in $\approx 220$~ms on a 32-core AMD EPYC CPU or $\approx 10.5$~ms on an NVIDIA H100 GPU for the baseline $N_b=5$ configuration. 
Latency scales super-linearly with $N_b$, while VRAM grows nearly linearly, increasing by roughly 0.65~GB per two additional sub-bands.
For example, increasing $N_b$ from 3 to 9 raises latency $\sim6\times$ (from 5.3~ms to 31.2~ms) and memory usage $\sim2.5\times$ (from 1.3~GB to 3.2~GB). 
All configurations fit within a single commodity GPU (4~GB VRAM), enabling near-real-time use. 

\section{Related Work}
\label{sec:related}
Neural network-based channel estimation has explored several architectures. 
For instance, \ac{CNN}-based methods~\cite{soltani2019deep, ye2018power} interpolate sparse pilot measurements across the time-frequency grid, but cannot recover entire missing bands. 
\Ac{LSTM} and \ac{GRU} approaches~\cite{jiang2019neural} capture per-frequency temporal correlation, but process subcarriers independently, missing cross-band spectral context. 
Attention-based estimators are gaining interest: Channelformer~\cite{luan2023channelformer} uses an encoder-decoder with multi-head self-attention for OFDM DL estimation, improving pilot feature robustness under 5G NR.
Jiang \emph{et al.}~\cite{jiang2022accurate} apply a Transformer to temporal \ac{CSI} sequences for Doppler-resilient prediction.
However, both approaches are suboptimal as they either incur quadratic complexity or lose cross-axis context.

Inspired by axial attention~\cite{ho2019axial}, our method attends along both frequency and time, capturing full 2D context at $\mathcal{O}(TF^2+FT^2)$ rather than $\mathcal{O}((TF)^2)$. 
Closest prior work is CsiNet~\cite{wen2018deep}, where a deep autoencoder for massive MIMO downlink \ac{CSI} is used for feedback reduction from complete observations.
Our \emph{CFRTransformer} instead recovers interference-corrupted spectra where whole sub-bands may be missing for multiple snapshots. 
The use of complex-valued layers follows Trabelsi \emph{et al.}~\cite{trabelsi2017deep}, extending Cauchy-Riemann weight coupling to Transformer embeddings and output layers.
Physics-informed losses in channel learning are surveyed in \cite{rao2020physics,javid2025physics}, but combining \ac{PDP} fidelity, \ac{CIR} sparsity, and temporal smoothness in one differentiable objective is novel. 
Finally, while \ac{DTMC} models have been used for spectrum occupancy~\cite{kim2008efficient}, coupling them with a neural reconstruction pipeline and jointly randomizing occupancy and velocity during training is a new contribution.


\section*{Acknowledgment}

The work was supported in part by the xG-RIC project funded by the Federal Ministry of Research, Technology, and Space (BMFTR) under grant number 16KIS2429K.

\section{Conclusion}
\label{sec:conclusion}

We have presented \emph{CFRTransformer}, a physics-informed complex Transformer for reconstructing wideband \ac{CFR} from fragmented multi-band observations under Markov interference.  
The architecture tackles the specific challenges of wideband wireless channels: complex-valued data, 2D time-frequency structure, and Doppler-limited temporal coherence; through a holomorphic embedding layer, factored frequency-and-time self-attention, and frequency positional encoding.  

Training with a composite physics-informed loss and per-sample velocity randomization allows the model to generalize across the full pedestrian-to-vehicular mobility range.  
Evaluation against three classical baselines, over interference-occupancy and velocity sweeps demonstrates consistent improvements in mean \ac{PDP} similarity.  
An ablation study confirms that velocity-randomized training ensures high-velocity robustness, without compromising performance.  
Future work will extend the evaluation of \emph{CFRTransformer} to measured channel datasets from 5G~NR and 802.11 WiFi field trials.


\bibliographystyle{IEEEtran}
\bibliography{references}

@article{wei2022toward,
    author = {Wei, Zhongxiang and Liu, Fan and Masouros, Christos and Su, Nanchi and Petropulu, Athina P.},
    doi = {10.1109/MCOM.002.2100972},
    title = {{Toward Multi-Functional 6G Wireless Networks: Integrating Sensing, Communication, and Security}},
    pages = {65--71},
    journal = {IEEE Commun. Mag.},
    issn = {0163-6804},
    publisher = {IEEE},
    number = {4},
    volume = {60},
    year = {2022}
}

@article{soltani2019deep,
    author = {Soltani, Mohammad Dehghani and Pourahmadi, Vahid and Mirzaei, Ali and Sheikhzadeh, Hamid},
    doi = {10.1109/LCOMM.2019.2898944},
    title = {{Deep Learning-Based Channel Estimation}},
    pages = {652--655},
    journal = {IEEE Commun. Lett.},
    issn = {1558-2558},
    publisher = {IEEE},
    month = {4},
    number = {4},
    volume = {23},
    year = {2019}
}

@article{jiang2019neural,
    author = {Jiang, Wei and Schotten, Hans},
    doi = {10.1109/ACCESS.2019.2937588},
    title = {{Neural Network-Based Fading Channel Prediction: A Comprehensive Overview}},
    pages = {118112 -- 118124},
    journal = {IEEE Access},
    issn = {2169-3536},
    publisher = {IEEE},
    month = {8},
    volume = {7},
    year = {2019}
}

@inproceedings{vaswani2017attention,
    author = {Vaswani, Ashish and Shazeer, Noam and Parmar, Niki and Uszkoreit, Jakob and Jones, Llion and Gomez, Aidan N. and Kaiser, Lukasz and Polosukhin, Illia},
    title = {{Attention is all you need}},
    pages = {6000--6010},
    publisher = {Curran Associates Inc.},
    isbn = {978-1-5108-6096-4},
    address = {Long Beach, CA},
    booktitle = {Int. Conf. Neural Information Process. Sys.},
    month = {12},
    year = {2017}
}

@article{he2020modeldriven,
    author = {He, Hengtao and Wen, Chao-Kai and Jin, Shi and Li, Geoffrey Ye},
    doi = {10.1109/tsp.2020.2976585},
    title = {{Model-Driven Deep Learning for MIMO Detection}},
    pages = {1702--1715},
    journal = {IEEE Trans. Signal Process.},
    issn = {1053-587X},
    publisher = {IEEE},
    month = {2},
    volume = {66},
    year = {2020}
}

@article{clarke1968statistical,
    author = {Clarke, Richard Hedley},
    title = {{A statistical theory of mobile-radio reception}},
    pages = {957--1000},
    journal = {Bell System Technical Journal},
    publisher = {Bell Telephone Laboratories},
    number = {6},
    volume = {47},
    year = {1968}
}

@article{ye2018power,
    author = {Ye, H. and Li, G. Y. and Juang, B.},
    doi = {10.1109/LWC.2017.2757490},
    title = {{Power of Deep Learning for Channel Estimation and Signal Detection in OFDM Systems}},
    pages = {114--117},
    journal = {IEEE Wireless Commun. Lett.},
    issn = {2162-2337},
    publisher = {IEEE},
    month = {2},
    number = {1},
    volume = {7},
    year = {2018}
}

@article{kim2008efficient,
    author = {Kim, Hyoil and Shin, Kang G.},
    doi = {10.1109/TMC.2007.70751},
    title = {{Efficient Discovery of Spectrum Opportunities with MAC-Layer Sensing in Cognitive Radio Networks}},
    pages = {533--545},
    journal = {IEEE Transactions on Mobile Computing},
    issn = {1536-1233},
    publisher = {IEEE},
    number = {5},
    volume = {7},
    year = {2008}
}

@book{tse2005fundamentals,
    author = {Tse, David and Viswanath, Pramod},
    doi = {10.1017/CBO9780511807213},
    title = {{Fundamentals of Wireless Communication}},
    publisher = {Cambridge University Press},
    year = {2005}
}

@techreport{3gpp38901,
    author = {{-}},
    title = {{Study on channel model for frequencies from 0.5 to 100 GHz}},
    institution = {3rd Generation Partnership Project (3GPP)},
    location = {Sophia Antipolis, France},
    month = {11},
    type = {TR},
    year = {2020}
}

@article{luan2023channelformer,
    author = {Luan, Dianxin and Thompson, John S.},
    doi = {10.1109/TWC.2023.3244484},
    title = {{Channelformer: Attention Based Neural Solution for Wireless Channel Estimation and Effective Online Training}},
    pages = {6562--6577},
    journal = {IEEE Trans. Wireless Commun.},
    issn = {1558-2248},
    publisher = {IEEE},
    month = {10},
    number = {10},
    volume = {22},
    year = {2023}
}

@article{jiang2022accurate,
    author = {Jiang, Hao and Cui, Mingyao and Ng, Derrick Wing Kwan and Dai, Linglong},
    doi = {10.1109/JSAC.2022.3191334},
    title = {{Accurate Channel Prediction Based on Transformer: Making Mobility Negligible}},
    pages = {2717--2732},
    journal = {IEEE J. Sel. Areas Commun.},
    issn = {0733-8716},
    publisher = {IEEE},
    number = {9},
    volume = {40},
    year = {2022}
}

@article{xie2019precise,
	author = {Xie, Yaxiong and Li, Zhenjiang and Li, Mo},
	title = {{Precise Power Delay Profiling with Commodity Wi-Fi}},
	doi = {10.1109/TMC.2018.2860991},
	issn = {1536-1233},
    journal = {IEEE Trans. Mob. Comput.},
    month = Jun,
	number = {6},
	pages = {1342--1355},
	publisher = {IEEE},
	volume = {18},
	year = {2019},
}

@article{rao2020physics,
    author = {Rao, Chengping and Sun, Hao and Liu, Yang},
    doi = {10.1016/j.taml.2020.01.039},
    title = {{Physics-informed deep learning for incompressible laminar flows}},
    pages = {207--212},
    journal = {Theor. Appl. Mech. Lett.},
    publisher = {Elsevier},
    number = {3},
    volume = {10},
    year = {2020}
}

@article{wen2018deep,
    author = {Wen, Chao-Kai and Shih, Wan-Ting and Jin, Shi},
    doi = {10.1109/LWC.2018.2818160},
    title = {{Deep Learning for Massive MIMO CSI Feedback}},
    pages = {748--751},
    journal = {IEEE Wireless Commun. Lett.},
    issn = {2162-2337},
    publisher = {Institute of Electrical and Electronics Engineers (IEEE)},
    number = {5},
    volume = {7},
    year = {2018}
}

@inproceedings{javid2025physics,
    author = {Javid, Alireza and Gonz{\'{a}}lez-Prelcic, Nuria},
    title = {{Physics-Informed Neural Networks for Wireless Channel Estimation with Limited Pilot Signals}},
    address = {San Diego, CA},
    booktitle = {AI and ML for Next-Gen. Wireless Commun. Netw.},
    month = {9},
    year = {2025}
}

@techreport{ho2019axial,
    author = {Ho, Jonathan and Kalchbrenner, Nal and Weissenborn, Dirk and Saliman, Tim},
    doi = {10.48550/arXiv.1912.12180},
    title = {{Axial attention in multidimensional transformers}},
    institution = {arXiv},
    month = {12},
    type = {cs.CV},
    year = {2019}
}

@techreport{trabelsi2017deep,
    author = {Trabelsi, Chiheb and Bilaniuk, Olexa and Zhang, Ying and Serdyuk, Dmitriy and Subramanian, Sandeep and Santos, Jo{\~{a}}o Felipe and Mehri, Soroush and Rostamzadeh, Negar and Bengio, Yoshua and Pal, Christopher J},
    doi = {10.48550/arXiv.1705.09792},
    title = {{Deep complex networks}},
    institution = {arXiv},
    month = {5},
    type = {cs.NE},
    year = {2017}
}

\end{document}